\documentclass[conference,a4paper]{IEEEtran}
\IEEEoverridecommandlockouts

\usepackage{algorithm,algorithmic,amsmath,amssymb,amsthm,array,bibentry,cite,color,comment,enumerate,enumitem,eurosym,float,graphicx}
\usepackage{lettrine,mathrsfs,multirow,nomencl,pict2e,psfrag,ragged2e,setspace,subfigure,supertabular,tabularx,url}
\usepackage[english]{babel}
\usepackage[absolute]{textpos}

{		}
{ 		}
{ 		\newtheorem{lemma}{Lemma}}
{ 		}
{		
{ 		}
{		
{		}
{		}
{ 		\newtheorem*{corollary}{Corollary}}
{ 		}
{		

\newcommand{\s}{\mathbf{s}}

\newcommand{\w}{\mathbf{w}}

\newcommand{\y}{\mathbf{y}}

\newcommand{\setC}{\mathcal{C}}

\newcommand{\setR}{\mathcal{R}}
\newcommand{\setS}{\mathcal{S}}

\newcommand{\snr}{\mathrm{snr}}
\newcommand{\mse}{\mathrm{mse}}

\DeclareMathOperator*{\ex}{\mathsf{E}}
\newcommand{\bydef}{\stackrel{\cdot}{=}}
\newcommand{\prob}{\mathsf{P}\!}
\newcommand{\var}{\mathsf{Var}\!}

\title{Single-Pole IIR Channel Power Prediction with Variable Delays}
\author{\IEEEauthorblockN{Jes\'{u}s Arnau}
\IEEEauthorblockA{Mathematical and Algorithmic Sciences Lab \\ France Research Center, Huawei Technologies Co. Ltd. \\
				 20 Quai du Point du Jour, 92100 Boulogne-Billancourt, France. \\
						  Email: jesus.arnau@huawei.com}}
\makeindex

\begin{document}
\begin{textblock}{14}(1,.5)
\begin{center}
Paper presented at IEEE GLOBECOM 2015, San Diego, California.
\end{center}
\end{textblock}

\maketitle

\begin{abstract}
Exploiting outdated channel quality indicators is crucial in most adaptive wireless communication systems. This is often done through channel prediction based on previous received indicators. In this paper, we analyze the case where the feedback delay experienced by the quality indicators is not constant, but random. Focusing on a single-pole IIR predictor, we obtain analytical expressions for the MSE and the filter parameters, and study the throughput behavior through Monte Carlo simulations. Results show that prediction provides a performance advantage for average delays smaller than $30$\,ms for low terminal speeds.
\end{abstract}


\section{Introduction} \label{sec:intro}
In wireless communication systems using adaptive coding and  modulation (ACM), dealing with outdated channel quality indicators (CQI) poses a relevant challenge. Consider the following uplink example: in order to select the appropriate modulation and coding scheme (MCS), the user equipment (UE) receives an indicator from the base station (BS) stating the quality of the channel. Because of the delays involved, such indicator will be outdated at the moment of using it, a phenomenon sometimes called {\em CQI aging} \cite{Akl2012} .

Apart from permanent delays given by BS processing and propagation, a crucial delay contribution is given by the time difference with the last channel estimation. For example, the LTE uplink uses periodic sounding reference signals (SRS) to estimate the channel, and so their period will determine how outdated the channel estimates are at the moment of using them.

Another possibility is to extract the CQI only from previous data transmissions. This uses fewer signaling resources, but introduces a new problem: the delays involved are now {\em random}, as a consequence of the packet arrival instants being random themselves. It shall be remarked that this is specially relevant for machine to machine (M2M) communications. In such systems, terminals use the channel sparingly and to transmit short messages, and superfluous signaling is kept to a minimum.

In this work, we compare the performance of channel prediction under fixed and random feedback delays. For its simplicity, and competitive performance, we illustrate the case of the single-pole IIR predictor in \cite{Cui2010}. We obtain closed-form expressions for the prediction mean-squared error (MSE), and for the optimum filter parameters. These will be complemented by Monte Carlo simulations that showcase throughput under realistic conditions. Results will show the different trends present with both types of delays. Moreover, they will illustrate the tipping points above which prediction offers little performance advantage.

The need for channel prediction to compensate for CQI aging has been longly recognized in the literature (see e.g. \cite{Duel-Hallen2007,Svensson2007}, and references therein). Because of the numerous trade-offs involved, the topic has kept attracting attention in the last years \cite{Akl2012, Bae2013, Ni2013, Abdallah2014}. To the best of our knowledge, existing works do not specifically address random feedback delays.

The remainder of the document is structured as follows: Section~\ref{sec:sys_model} explains the system model and assumptions, Section~\ref{sec:analysis} collects the analytical derivations of the MSE for fixed and variable delays, Section~\ref{sec:sim_results} reports some simulation results, and finally conclusions are summarized in Section~\ref{sec:conclusions}.

\section{System model} \label{sec:sys_model}
\subsection{Signal model}\label{subsec:signal_model}
The signal model considered is given by
\begin{equation}
\y_\ell = \sqrt{\snr}\cdot h_\ell \s_\ell+\w_\ell
\end{equation}
where $\s_\ell$ denotes the $\ell$-th block of $B$ independent, unit-power transmitted symbols, $\s_\ell = [s_\ell, s_{\ell+1}, \dotsc, s_{\ell+B}]^T$; $\y_\ell$ denotes the received block, and $\w_\ell$ contains $B$ zero-mean, unit-power samples accounting for Gaussian noise plus interference.

We assume that each transmitted block undergoes a single channel realization ({\it block fading}), and that blocks are transmitted using a finite set of $M$ modulation and coding schemes, $\setC  = \{C_1, C_2, \cdots, C_M\}$, characterized by their associated signal-to-interference-plus noise (SINR) thresholds $\setS$ and spectral efficiencies $\setR$. We will further assume that the $\ell$-th block, using the $j$-th MCS, is transmitted successfully if and only if $\snr|h_\ell|^2>S_j$.

The channel represents a Rayleigh fading scenario, so that $h_\ell\sim\mathcal{CN}(0,1)$; time variations in the fading process are ruled by the Doppler spectrum, whose bandwidth is given by $f_\mathrm{d} = v/cf_\mathrm{c}$; $v$ is the terminal speed, $c$ is the speed of light ($3\cdot10^8$\,m/s) and $f_\mathrm{c}$ is the carrier frequency. We will go deeper into the details of the channel's time correlation in the next section.

For future reference, we should note that each power sample $|h_\ell|^2$ follows a scaled Chi-squared distribution with two degrees of freedom, $2|h|^2\sim\chi^2(2)$; therefore, the following identities hold: $\ex\left[|h|^2\right] = 1$, $\ex\left[|h|^4\right] = 2.$

\subsection{Traffic pattern}\label{subsec:traffic}
As explained, we will allow the transmission time instants to be random. This means that the set $h_\ell$ will not be a periodic sampling of a continuous-time channel process: instead, the underlying channel will be sampled at random time instants, and thus the correlation between adjacent samples will vary.

Let us assume that arrivals follow a Poisson process. Then, the time interval between two consecutive channel samples, $h_\ell$ and $h_{\ell+1}$, denoted as $\tau_\ell$, will be exponentially distributed
\begin{equation}\label{eq:exp_arrivals}
\tau_\ell\sim\mathrm{Exp}(1/\overline{T})
\end{equation}
where $\overline{T}$ denotes the mean time between blocks. We shall remark that such an exponential process is memoryless, and that, as suggested by the notation in (\ref{eq:exp_arrivals}), the random variables $\tau_\ell$ are i.i.d; we will often drop the subindex when unnecessary.

\subsection{Time correlation}\label{subsec:time_correlation}
The classical Jakes-Clarke model states that, if $h_\ell$ is a Rayleigh fading process, then $\ex\left[h_\ell h_{\ell+L}^*\right] = J_0\left(2\pi f_\mathrm{d}\tau|L-\ell|\right)$. However, the presence of the Bessel function makes it difficult to analyze this model, and specially in our case where $\tau$ is not a constant.

For the rest of the paper, we will assume a simplified correlation model, where $h_\ell$ is a first-order autoregressive process:
\begin{equation}\label{eq:gm}
h_{\ell+1} = \rho h_\ell+\sqrt{1-\rho^2}\cdot e_{\ell+1}
\end{equation}
where the $e_{\ell}$ are independent complex standard normal random variables, and $\rho$ is the correlation factor between adjacent  samples:
\begin{equation}
\rho\bydef\ex\left[h_\ell h_{\ell+1}^*\right].
\end{equation} 

\subsubsection{Evenly spaced arrivals}\label{sec:evenly_spaced}
If, as traditionally, channel samples are evenly spaced over time, then the correlation coefficient between them will be constant. It trivially follows that
\begin{equation}
\begin{split}
\ex\left[h_{\ell+L}h_{\ell}^*\right] = \rho^L.
\end{split}
\end{equation}

\subsubsection{Randomly spaced arrivals}\label{sec:variably_spaced}
The case where arrivals are randomly spaced is more difficult to handle (even in the case of i.i.d intervals from Section~\ref{subsec:traffic}). To tackle it, we propose the following model with random correlation coefficients:
\begin{equation}
\begin{split}
h_{\ell+1} &= \rho_\ell h_\ell+\sqrt{1-\rho_\ell^2} e_{\ell+1}. \\
\end{split}
\end{equation}
Here, each $\rho_\ell$ is random, and given by
\begin{equation}
\rho_\ell  = J_0(2\pi f_\mathrm{d}\tau_\ell).
\end{equation}
We should recall that $\tau_\ell$ are exponentially-distributed random variables.

Correlation between any two channel samples is now given by
\begin{equation}\label{eq:variable_corr}
\ex_h\left[h_{\ell+L}h_{\ell}^*\right] = \prod_{i=1}^L\rho_{\ell+i},
\end{equation}
which again is a random variable and depends on the arrivals process. Thus, it will be useful to define the {\em average correlation coefficient}:
\begin{equation}
\overline\rho \bydef \ex_\tau\left[\rho\right] = \ex_\tau\left[\ex_h\left[h_\ell h_{\ell+1}^*\right]\right].
\end{equation}

\subsection{Method under study}\label{subsec:methods}
Let $\gamma_\ell$ denote the $\ell$-th realization of the process we want to predict, then the single-pole IIR predictor proposed in \cite{Cui2010} is given by
\begin{equation}\label{eq:iir}
\hat\gamma_{\ell+1} = (1-\alpha)\hat\gamma_\ell+\alpha\gamma_\ell.
\end{equation}


For future usage, note that the mean squared error of the prediction is given by \cite[Eq.~13]{Cui2010}
\begin{equation}\label{eq:mse_cui}
\begin{split}
\mse &\bydef \ex\left[|\hat\gamma_{\ell+1}-\gamma_{\ell+1}|^2\right] \\
	  &= \frac{2}{2-\alpha}\left(\ex\left[|\gamma|^2\right]-\alpha\sum_{i=1}^\infty(1-\alpha)^{i-1}\ex\left[\gamma_{\ell}\gamma_{\ell-i}^*\right]\right).
\end{split}
\end{equation}

Before going further, we should note that (\ref{eq:iir}) admits the following expression in terms of the infinite impulse response of the prediction filter, denoted below by $g$:
\begin{equation}\label{eq:inf_sum}
\begin{split}
\hat\gamma_{\ell+1} &= \sum_{i=0}^\infty g_i\gamma_{\ell-i} \\
					&=\alpha\sum_{i=0}^\infty(1-\alpha)^i\gamma_{\ell-i}.
\end{split}
\end{equation}

\section{Analysis of single-pole IIR prediction}\label{sec:analysis}
In this section, we analyze the performance of the predictor above in terms of MSE. We will start by comparing two different approaches to power prediction, and choose one of them for the rest of the paper. We will analytically obtain the MSE as a function of $\alpha$, both with fixed and random block spacing, and also the expression of the optimum $\alpha$.

\subsection{Power vs. amplitude estimation}\label{subsec:power_vs_amplitude}
As explained in Section~\ref{subsec:signal_model}, each block is transmitted using an MCS whose transmission success will depend on the instantaneous SINR it experiences. Consequently, we will be interested in predicting the instantaneous channel power, given by $\snr|h_{\ell+1}|^2$. We will briefly analyze the behavior of two different power prediction alternatives.

Assume that we have a prediction of the future channel sample $h_{\ell+1}$, and that we want to derive the instantaneous power from it. This predictor is inherently biased:
\begin{equation}
\begin{split}
\ex\left[|\hat h_{\ell+1}|^2\right] 	&= \ex\left[\left|\alpha\sum_{i=0}^\infty(1-\alpha)^i h_{\ell-i}\right|^2\right] \\
									&=\frac{\alpha}{2-\alpha}\ex\left[|h|^2\right]\\ &+2\alpha^2\sum_{j=0}^\infty\sum_{k=0}^j(1-\alpha)^{j+k}\ex\left[h_{\ell-j}h_{\ell-k}^*\right] \\
									&\neq\ex\left[|h|^2\right]
\end{split}
\end{equation}
unless $\alpha=1$, in which case we would be simply using the previous sample as a predictor.

On the other hand, prediction applied directly over power estimates is naturally unbiased. For the rest of this work, we will assume that we operate directly over power samples. For simplicity, we will assume perfect knowledge of the previous power sample $\gamma_\ell = \snr|h_\ell|^2$, and focus on the uncertainty given by the time dynamics of $h_\ell$. Further discussions on the bias can be found in \cite{Ekman2002}.

\subsection{Special cases of $\alpha$}

\subsubsection{Small $\alpha$ }\label{subsec:small_alpha}
When $\alpha<<1$ we can approximate the predicted sample as $\hat\gamma_{\ell+1}\approx\hat\gamma_{\ell} \approx \hat\gamma_{\ell-1} \cdots \approx \hat\gamma_{-\infty}$, where $\hat\gamma_{-\infty}$ is the first prediction guess used to initialize the algorithm. Then, the MSE
\begin{equation}
\mse\approx\ex\left[|\hat\gamma_{\ell+1}-\gamma_{-\infty}|^2\right]
\end{equation}
is minimized when $\hat\gamma_{-\infty} = \ex\left[\gamma_\ell\right]$, and in consequence 
\begin{equation}
\mse = \var\left[\gamma_\ell\right]
\end{equation}
by the definition of variance.

In summary, setting a very low $\alpha$ ends up giving a constant value as a prediction, and its lowest MSE is given by the variance of the process.

\subsubsection{Almost-1 $\alpha$ }\label{susubbsec:alpha_one}
When $\alpha$ is almost equal to one, then $\hat\gamma_{\ell+1}\approx\gamma_\ell$, and the MSE is given by
\begin{equation}
\begin{split}
	\mse &\approx \ex\left[|\gamma_\ell-\gamma_{\ell+1}|^2\right] \\
		 &=2\left(\ex\left[|\gamma_\ell|^2\right]-\ex\left[\gamma_\ell\gamma_{\ell+1}^*\right]\right).
\end{split}
\end{equation}
The predictor is the previous sample. The error decreases as correlation increases; in the limit where $\gamma_{\ell+1} \approx \gamma_\ell$ (constant channel), then $\mse=0$.

\subsection{Fixed inter-block period}\label{subsec:fixed}
Let us start by assuming that data is transmitted at a constant rate (or that there exists a periodic probing of the channel, as discussed in Section~\ref{sec:intro}), so that the time distance between consecutive blocks is always the same. In this case, $\tau_\ell = T_\mathrm{s}\ \forall\ell$, and in consequence $\rho_\ell = \rho =J_0\left(2\pi f_\mathrm{d}T_\mathrm{s}\right) $. The MSE is given in closed form by the following lemma.
\begin{lemma}
With fixed inter-block period, the MSE of the single-pole predictor is given by
%
\begin{equation}\label{eq:mse_gm}
\mse =     \frac{2\snr^2}{2-\alpha}\left(1-\frac{\alpha}{\alpha-1+\rho^{-2}}\right), \quad 0<\alpha<2.
\end{equation}
\end{lemma}
\begin{IEEEproof}
See Appendix~\ref{app:proof_mse}.
\end{IEEEproof}

Note that, if $\alpha$ approaches zero without reaching it, then the $\mse$ approaches $\snr^2$, which is the variance of $\snr|h_\ell|^2$ as anticipated in Section~\ref{subsec:small_alpha}.

\begin{corollary}
The optimum value of $\alpha$ is given by
\begin{equation}\label{eq:alpha_opt_fixed}
\begin{split}
\alpha_\mathrm{opt} &= \frac{1}{2}\left(3-\rho^{-2}\right) \\
					&=\frac{1}{2}\left(3-J_0\left(2\pi f_\mathrm{d}T_{s}\right)^{-2}\right).
\end{split}
\end{equation}
\end{corollary}
\begin{IEEEproof}
Differentiating (\ref{eq:mse_gm}) with respect to $\alpha$, we obtain
\begin{equation}
\frac{\partial\mse}{\partial\alpha} = \frac{2\snr^2}{(2-\alpha)^2}\frac{(\rho^2-1)\left(\rho^2(2\alpha-3)+1\right)}{\left(1+(\alpha-1)\rho^2\right)^2}.
\end{equation}
Equating to zero and solving for $\alpha$ gives the result.
\end{IEEEproof}

When the argument of the Bessel function becomes small, then $\rho$ tends to $1$ and the optimum $\alpha$ tends to $1$ as well, which is consistent with the analysis in Section~\ref{susubbsec:alpha_one}: when correlation is high, all the importance is given to the previous sample. On the other hand, when the argument grows large and $\rho$ tends to zero, the optimum value of $\alpha$ would tend to a large negative quantity. This, however, cannot be allowed, as $\alpha$ has to be between $0$ and $2$ for convergence; we shall clip the negative values to zero when needed. This means that, when correlation is very low, the channel will be predicted always by the same sample, the one used to initialize the algorithm.

We shall remark that $\alpha_\mathrm{opt}$ will be between $0$ and $1$, but beware that this conclusion relies on our assumptions: other correlation models could potentially yield optimum values above $1$.

\subsection{Random inter-block period}\label{subsec:variable}
For the random inter-block period, we need to average over both the realizations of $h_\ell$ and $\tau_\ell$. Doing so, the MSE reads as
\begin{equation}
\begin{split}
\mse 	&= \ex_\tau\left[\ex_h\left[|\hat\gamma_{\ell+1}-\gamma_{\ell+1}|^2\right]\right],
\end{split}
\end{equation}
and its expression is given in the following lemma.

\begin{lemma}
The MSE of the single-pole predictor with exponential inter-block delay is given by
\begin{equation}\label{eq:mse_variable}
\mse = \frac{2\snr^2}{2-\alpha}\left(1-\frac{\alpha}{\alpha-1+\frac{\pi}{2}K(-16\pi^2f_\mathrm{d}^2\overline T^2)^{-1}}\right)
\end{equation}
where $K(x)$ denotes the {\em complete elliptic integral of the first kind} \cite[Sec.~17.3]{Abramowitz1964} and $\overline{T}$ is the average time between consecutive blocks.
\end{lemma}

\begin{IEEEproof}
See Appendix~\ref{app:proof_mse2}.
\end{IEEEproof}

\begin{corollary}
The value of $\alpha$ that minimizes the MSE is given by
\begin{equation}
\overline{\alpha_\mathrm{opt}} = \frac{1}{2}\left(3-\frac{\pi}{2}K(-16\pi^2f_\mathrm{d}^2\overline T^2)^{-1}\right).
\end{equation}
\end{corollary}

We will compare these expressions numerically in the following section.
\section{Simulation results}\label{sec:sim_results}

\subsection{Numerical evaluation of the MSE}\label{subsec:num_eval}
We first evaluate the MSE given by (\ref{eq:mse_gm}) and (\ref{eq:mse_variable}), as a function of $\alpha$, for different correlation levels.

\begin{figure}
\centering
\includegraphics[width=\columnwidth, clip=true, trim = 0 3 1 1]{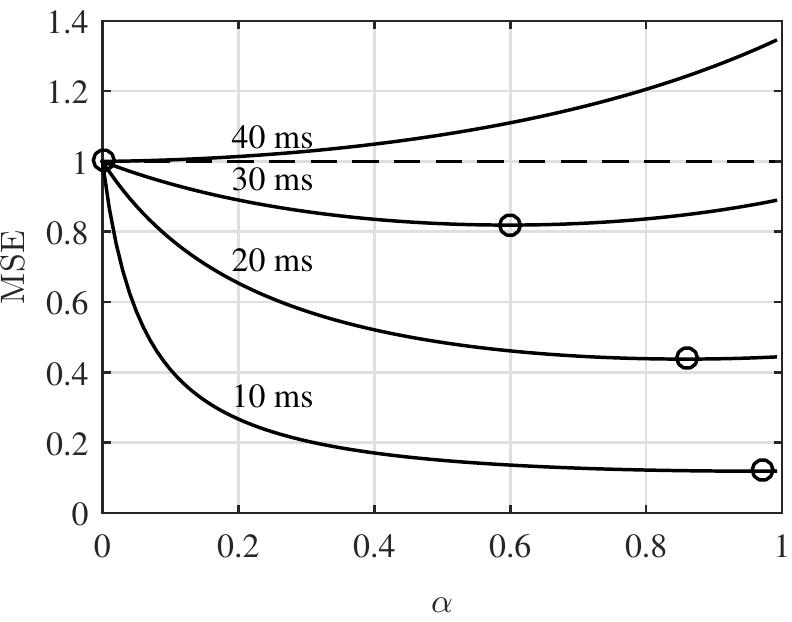}
\caption{MSE as a function of $\alpha$ for different values of $T_\mathrm{s}$; the dashed line shows the variance, and squares mark the minimum MSE.} 
\label{fig:mse_gm}
\end{figure}

Figure~\ref{fig:mse_gm} shows the results with fixed block separation $T_\mathrm{s} \in \{10, 20, 30, 40\}$\,ms. The speed $v$  is set to $3$\,km/h, the carrier frequency $f_\mathrm{c}$ to $2$\,GHz, and the SNR to $0$\,dB; the dashed line is the variance, which would be the MSE obtained by always using the mean as a prediction. We can see that, for delays above $40$\,m/s, there is no gain in terms of MSE with respect to just using the mean; furthermore, the gains with a delay of $30$\,m/s would be reduced to at most $20$\%. 

\begin{figure}
\centering
\includegraphics[width=\columnwidth, clip=true, trim = 0 3 1 1]{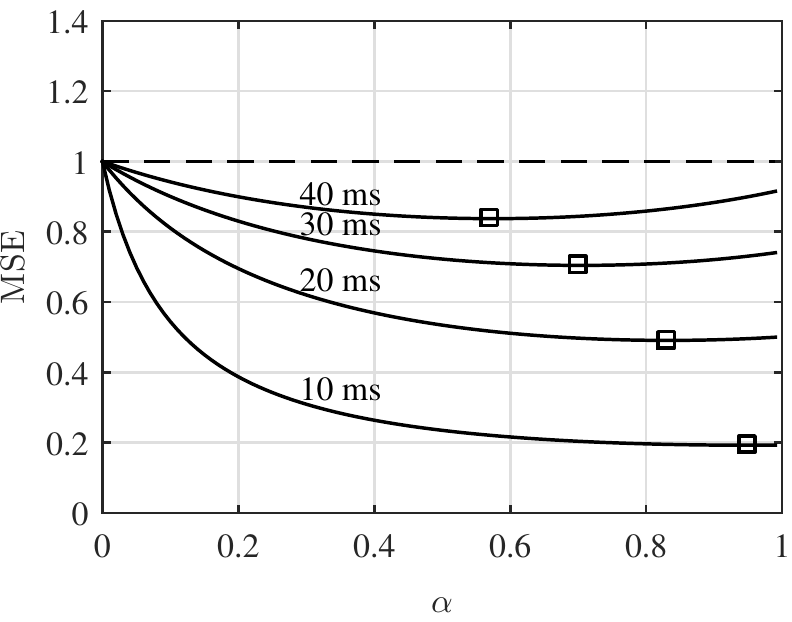}
\caption{MSE as a function of $\alpha$ for different values of $\overline{T}$; the dashed line shows the variance, and circles mark the minimum MSE.} 
\label{fig:mse_gm2}
\end{figure}

Similarly, Figure~\ref{fig:mse_gm2} evaluates the expressions for the variable separation case with  $\overline{T}\in \{10, 20, 30, 40\}$\,ms. Here, for a mean block separation of $40$\,m/s, there is still almost $20$\% advantage in terms of MSE when doing prediction. This is because a sufficient number of samples of the process $\tau_\ell$ will take values well below the mean, making prediction useful.

\begin{figure}
\centering
\includegraphics[width=\columnwidth, clip=true, trim = 0 0 0 0]{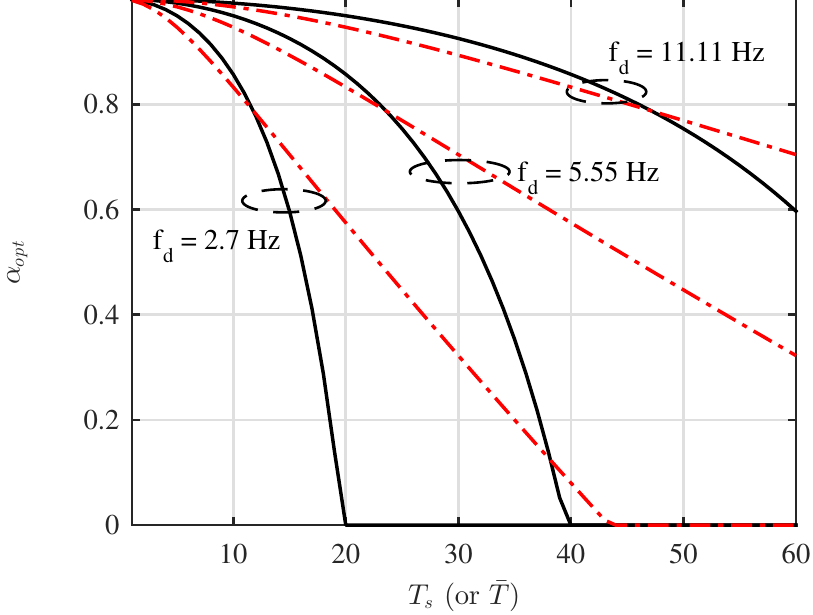}
\caption{$\alpha$ as a function of $T_\mathrm{s}$ for the case of fixed intervals (solid lines), and as a function of $\overline{T}$ for variable intervals (dashed lines).} 
\label{fig:a_opt}
\end{figure}

Finally, we evaluate the optimum $\alpha$ for both cases in Figure~\ref{fig:a_opt} for different speeds (represented on the plot by different values of $f_\mathrm{d}$). It is worth noticing the sensitivity of the fixed interval case: a slight deviation in $T_\mathrm{s}$ causes an abrupt change in $\alpha_\mathrm{opt}$, whereas significant deviations are needed in $\overline{T}$ to see the same effect.

\subsection{Throughput simulation}\label{subsec:thrp_sim}
We next report average throughput (average over time) results through Monte Carlo simulations. Unless otherwise stated, the simulation parameters are the same as in the previous section. We use adaptive coding and modulation, and select the MCS based on the predicted value. For comparison, we also consider the following techniques:
\begin{itemize}
\item {\em Perfect prediction:} as an upper bound, we show the average throughput that would be obtained with perfect knowledge of the future channel state.
\item {\em Previous sample:} we test the case of simply using the previous channel sample as a predictor (equivalent to fixing $\alpha=1$).
\item {\em Fixed rate:} The same MCS $C_{j^*}$ is used in every transmission. It is selected as 
\begin{equation}
\begin{split}
j^* 
	&= \arg\max_j \ \left( 1-\prob\left[\snr|h_\ell|^2 \leq S_j\right]\right)R_j \\
	&= \arg\max_j \  \exp\left({-\frac{S_j}{2\snr}}\right)R_j.
\end{split}
\end{equation}
\end{itemize}

\begin{figure}
\centering
\includegraphics[width=\columnwidth, clip=true, trim = 0 0 0 0]{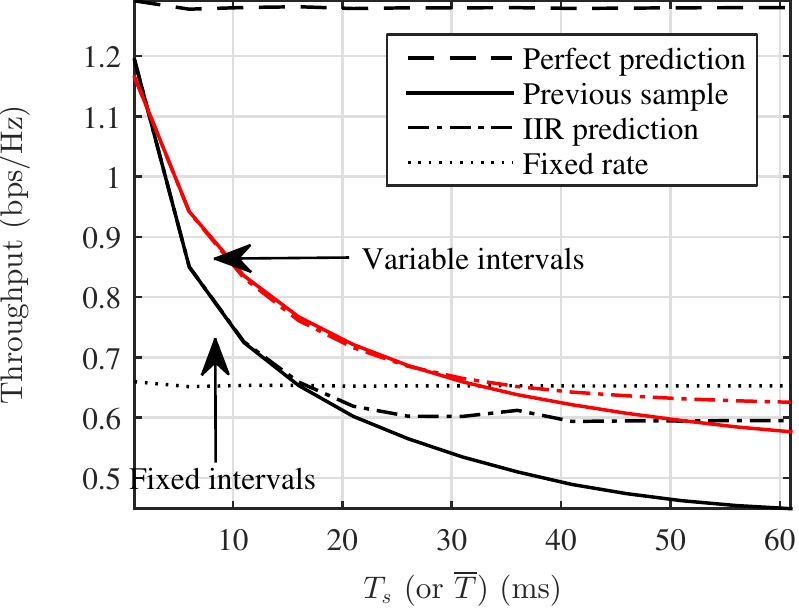}
\caption{Throughput as a function of $T_\mathrm{s}$ (fixed case) or $\overline{T}$ (variable case). $\snr = 5$\,dB.} 
\label{fig:tt}
\end{figure}

Results are shown in Figure~\ref{fig:tt} for $\snr = 5$\,dB; the optimum value of $\alpha$ is used in every case. We can see that, for both fixed and variable delays, using the previous sample as a predictor performs close to using the optimum $\alpha$ for low delays, and comparably worse for higher delays; the difference seems to be smaller in the random case. In either case, this kind of prediction offers little or no advantage for (average) delays above $30$\,m/s, even with a speed as low as $3$\,km/h.
\section{Conclusions}\label{sec:conclusions}
We have analyzed the performance of a channel power predictor with fixed and random delays. Closed-form expressions for the MSE and the predictor parameters have been obtained for both cases. Results show that, for low speeds, prediction provides an advantage for average delays up to about $30$\,ms. Extensions to more general predictors and channel models will be the subject of future work.

\appendices

\section{Analytical MSE with fixed intervals}\label{app:proof_mse}
Let us start from the fact that
\begin{equation}\label{eq:corrs}
 \ex_h\left[\gamma_\ell\gamma_{\ell+i}^*\right] = \snr^2\cdot\left(1+\ex_h\left[h_\ell h_{\ell+i}^*\right]^2\right),
\end{equation}
as shown among others in \cite[Eq. 14]{Wang2005}. Substituting (\ref{eq:corrs}) in (\ref{eq:mse_cui}) we obtain
\begin{equation}
\begin{split}
\mse   &= \frac{2}{2-\alpha}\left(\ex\left[\gamma_\ell^2\right]-\alpha\sum_{i=1}^\infty(1-\alpha)^{i-1}\right.\\ 
&\times \left. \vphantom{\sum_{i=1}^\infty}\snr^2\cdot\left(1+\ex\left[h_\ell h_{\ell+i}^*\right]^2\right)\right) \\
	   &=\frac{2\snr^2}{2-\alpha}\left(2-\alpha\sum_{i=1}^\infty(1-\alpha)^{i-1}\left(1+\rho^{2i}\right)\right) \\
	   &=\frac{2\snr^2}{2-\alpha}\left(2-\alpha\sum_{i=1}^\infty(1-\alpha)^{i-1}-\alpha\sum_{i=1}^\infty(1-\alpha)^{i-1}\rho^{2i}\right) \\
	   &=\frac{2\snr^2}{2-\alpha}\left(2-\alpha\sum_{i=1}^\infty(1-\alpha)^{i-1}-\frac{\alpha}{\alpha-1+\rho^{-2}}\right).
\end{split}
\end{equation}
Finally, noting that
\begin{equation}
\alpha\sum_{i=1}^\infty(1-\alpha)^{i-1} = \begin{cases}
										1 & \mathrm{if} \ \alpha=0 \\
										0 & \mathrm{if} \ 0<\alpha<2
										\end{cases}
\end{equation}
we finish the proof.  \IEEEQED
\section{Analytical MSE with random intervals}\label{app:proof_mse2}
If we average over both the channel realizations and arrival process, we obtain:
\begin{equation}
\begin{split}
\mse 	&= \ex_\tau\left[\ex_h\left[|\hat\gamma_{\ell+1}-\gamma_{\ell+1}|^2\right]\right] \\
		&=\ex_\tau\left[ \frac{2}{2-\alpha}\left(\ex\left[\gamma_\ell^2\right]-\alpha\sum_{i=1}^\infty(1-\alpha)^{i-1}\ex_h\left[\gamma_{\ell}\gamma_{\ell-i}^*\right]\right)\right] \\
		&= \frac{2\snr^2}{2-\alpha}\left(2-\alpha\sum_{i=1}^\infty(1-\alpha)^{i-1}\ex_\tau\left[1+\ex_h\left[h_{\ell}h_{\ell-i}^*\right]^2\right]\right)
\end{split}
\end{equation}
where we have used (\ref{eq:mse_cui}), together with (\ref{eq:corrs}). Now, plugging (\ref{eq:variable_corr}), we obtain
\begin{equation}
\begin{split}
\mse 	&=  \frac{2\snr^2}{2-\alpha}\left(2-\alpha\sum_{i=1}^\infty(1-\alpha)^{i-1}\left(1+\ex_{\tau}\left[ \prod_{j=1}^i\rho_{\ell-j}^2\right]\right)\right)\\
&=  \frac{2\snr^2}{2-\alpha}\left(2-\alpha\sum_{i=1}^\infty(1-\alpha)^{i-1}\left(1+ \prod_{j=1}^i\ex_{\tau}\left[\rho_{\ell-j}^2\right]\right)\right)\\
&=  \frac{2\snr^2}{2-\alpha}\left(2-\alpha\sum_{i=1}^\infty(1-\alpha)^{i-1}\left(1+ \ex_{\tau}\left[\rho^2\right]^i\right)\right)
\end{split}
\end{equation}
where we have used the fact that the $\rho_\ell$ are i.i.d random variables. Defining $\overline{\rho^2} \bydef \ex_{\tau}\left[\rho^2\right]$ and using (\ref{eq:mse_gm}), we arrive at
\begin{equation}\label{eq:mse_var}
 \mse = \frac{2\snr^2}{2-\alpha}\left(1-\frac{\alpha}{\alpha-1+\overline{\rho^2}^{-1}}\right)
\end{equation}
whenever $\alpha\neq0$.

It only remains to compute the expectation $\ex_{\tau}\left[\rho^2\right]$. Since $\tau$ is exponentially distributed: 
\begin{equation}
\begin{split}
\overline{\rho^2}	 &= \ex_{\tau}\left[\rho^{2}\right] \\
					 &= \frac{1}{\overline T}\int_0^\infty J_0(2\pi f_\mathrm{d}\tau)^2e^{-{\tau}/{\overline T}}\,\mathrm{d}\tau = \frac{2}{\pi}K\left(-16\pi^2f_\mathrm{d}^2\overline T^2\right)	
\end{split}
\end{equation}
which follows from \cite[Eq.~6.6112.4]{Gradshteyn2007}, and where $K(x)$ denotes the {\em complete elliptic integral of the first kind} \cite[Sec.~17.3]{Abramowitz1964}. The MSE finally reads as
\begin{equation}
\mse = \frac{2\snr^2}{2-\alpha}\left(1-\frac{\alpha}{\alpha-1+\frac{\pi}{2}K(-16\pi^2f_\mathrm{d}^2\overline T^2)^{-1}}\right)
\end{equation}
which finishes the proof. \IEEEQED

\bibliographystyle{IEEEtran}
\bibliography{IEEEabrv,acmbib,books_and_others}
\end{document}